\documentclass[journal]{IEEEtran}
\IEEEoverridecommandlockouts
\usepackage{cite}
\usepackage{amsmath,amssymb,amsfonts}
\usepackage{layout}
\usepackage{algorithm}
\usepackage{algorithmic}
\usepackage{graphicx}
\usepackage{textcomp}
\usepackage{xcolor}
\usepackage{amsmath}
\usepackage{multirow}
\usepackage{graphicx}
\usepackage{tabularx}
\usepackage{amsthm}
\usepackage{subfigure}
\usepackage{booktabs}

\newtheorem{note}{Note}

\usepackage{subfig}
\DeclareMathOperator{\sinc}{sinc}
\DeclareMathOperator{\var}{Var}
\DeclareMathOperator{\cov}{Cov}

\begin{document}

\title{A Phase-Balanced Differential Rydberg-Atom Quantum Receiver for Weak GNSS Signal Detection}
\author{A. K. Mandal, F. Zhang, J. H. Mikkelsen, and M. Shen\\
{\small Department of Electronic Systems, The Technical Faculty of IT \& Design, Aalborg University, 9220 Aalborg \O st, Denmark}
}
\maketitle
%
\begin{abstract}
This paper proposes a phase-balanced differential Rydberg atomic quantum receiver (DRAQR) for weak-signal recovery, using local-oscillator phase control to reinforce differential signal responses and a shared squeezed-reservoir model to correlate inter-cell quantum projection noise (QPN) and photon shot noise (PSN). An analytical model incorporating QPN, PSN, intrinsic thermal noise (ITN), and inter-cell correlations quantifies signal-to-noise ratio (SNR) gain, outage-constrained minimum detectable power, and bit-error rate (BER) for square $M$-QAM. For the evaluated settings, SNR gain reaches approximately $10$~dB as PSN correlation approaches unity, with BER decreasing as correlation increases. Under Rician fading, at a $0$-dB threshold SNR and maximum outage probability of $0.1$, the minimum detectable power decreases from $-146.95$~dBm at $K=1.155$ to $-159.98$~dBm at $K=2.4$. Analytical results are validated against Monte-Carlo simulations over $150{,}000$ data runs per point for receiver-performance plots.
\end{abstract}
%
%
\section{Introduction}\label{intro}
Weak radio-frequency (RF) signal recovery is challenging in satellite navigation, wireless communication, and remote sensing due to fading channel, propagation loss, and time-varying channels. Rydberg-atom electrometry enables optical RF sensing through field-sensitive, highly excited atomic states, with electromagnetically induced transparency (EIT) mapping atomic responses onto probe-light signals. Recent studies extend beyond field calibration to waveform reception, receiver dynamics, and communications performance \cite{Sandidge2024,GongTCOM2026,ChenTWC2026,CuiTWC2026}.
%
%

Rydberg-atom quantum receivers (RAQRs) use optical excitation and electromagnetically induced transparency (EIT) to convert RF-field interactions into probe-transmission changes. While early schemes relied on Autler--Townes splitting (ATS), phase-sensitive heterodyne and interferometric methods enable modulation-phase and time-varying waveform recovery across diverse frequency bands. Recent communication-oriented models characterize the incident-field-to-photodetector signal path, examining calibration, local oscillator (LO) configuration, and receiver bandwidth \cite{GongTCOM2026,ChenTWC2026,CuiTWC2026}. Atomic master-equation and linear time-invariant models further quantify responses to time-varying fields beyond steady-state analysis \cite{ZhuDaiTWC2026,MalvaniaTQE2026}.

In superheterodyne receivers, a strong LO mixes with the weak received field to generate a beat response at the frequency difference between them, enabling recovery of signal amplitude, phase, and frequency. Recent studies have modeled receiver performance for classical communication waveforms. A cavity-coupled receiver demonstrated reception of a geostationary-satellite beacon and modulated satellite signals, reporting a field sensitivity of $21$~nVcm$^{-1}$Hz$^{-1/2}$ and an $8$-dB signal-to-noise ratio (SNR) over a $10$-Hz bandwidth under the reported conditions \cite{GongTCOM2026,ChenTWC2026,LeiTQE2026}. These results provide motivation for the modeling of weak satellite signals using atomic response, LO phase, and in-band receiver noise.

Recent studies enhance RF coupling to atomic vapor or optical readout using resonant cavities, compact resonators, optimized laser-array geometry, Fabry--P\'erot cavities, and LO-port resonators. Multi-band designs exploit distinct atomic transitions and microwave mixing to extend spectral coverage \cite{Sandidge2024,MaoRes2024,WuLaser2025,wu2024enhancing,YangLO2025,CuiTWC2026}. Other works investigate calibration-free reception below $1$~MHz and joint optimization of sensitivity and bandwidth \cite{ChenTWC2026,WangTIM2026}, while frequency-comb processing addresses ambiguity in wideband RF-frequency estimation with RAQRs \cite{WangTriComb2026}. These studies identify field enhancement, atomic operating point, LO design, and temporal response as coupled determinants of receiver sensitivity and bandwidth.

Recent communication-system models consider quantum projection noise (QPN), photon shot noise (PSN), blackbody-radiation-induced fields, receiver electronics, and laser and detector fluctuations when evaluating SNR, capacity, and measured photocurrent \cite{ZhuDaiTWC2026,ChenTWC2026,CuiTWC2026,GongTCOM2026}. Thus, increased field response alone does not guarantee a proportional improvement in the minimum detectable signal. Spatial processing has expanded to receiving arrays and beamforming, with atomic-array models for directivity and satellite Doppler localization \cite{GongTVTDOA2026,HanTCOM2026}. These advances highlight phase alignment and inter-channel spatial relationships alongside single-cell sensitivity.

This work investigates a novel phase-balanced, two-cell differential RAQR (DRAQR) for the recovery of global navigation satellite systems (GNSS)-like weak-signal. The cells share probe and coupling fields and receive nearly identical waveforms; the LO phase compensates the inter-cell phase offset so the desired atomic responses combine differentially. QPN and PSN are modeled with separate inter-cell covariance terms to represent signal enhancement in the differential mode and correlated-noise suppression in the common mode. 

%
The key contributions of this work are as follows:
\begin{enumerate}
\item We formulate a phase-balanced, dual-cell RAQR, i.e., a DRAQR, for weak RF detection, using controlled LO phase and differential readout to reinforce the desired signal under shared RF illumination.
\item We model inter-cell noise correlations through a shared squeezed reservoir, accounting for reservoir squeezing, coupling, atomic decoherence, transfer efficiency, optical squeezing, and detection loss.
\item We derive a noise model incorporating QPN, PSN, intrinsic thermal noise (ITN), and inter-cell covariance, and quantify their impact on differential noise and SNR. 
\item Finally, we analyze outage-constrained minimum power and square Gray-coded $M$-QAM BER, validated by the Monte-Carlo (MC) simulation results.
\end{enumerate}
The significance of this research can be observed under Rician fading, at $\gamma_{\rm th}=0$~dB and $P_{\rm out}^{(\max)}\approx0.1$, where the $P_b^{(\min)}$ decreases by $\approx13.0$~dB, from $-146.95$~dBm at $K=1.155$ to $-159.98$~dBm at $K=2.4$. Analytical and MC estimates agree within $+0.005$~dB over $150{,}000$ data runs.
%

\textbf{Organization:} 
Section~\ref{draqrArch} shows the DRAQR architecture, Section~\ref{pbDraqr} presents the DRAQR analyses, followed by results and conclusions in Section~\ref{res} and Section~\ref{conc}, respectively.

\section{DRAQR Architecture}\label{draqrArch}
\begin{figure}[t!]
\centering
\includegraphics[width=3.49in]{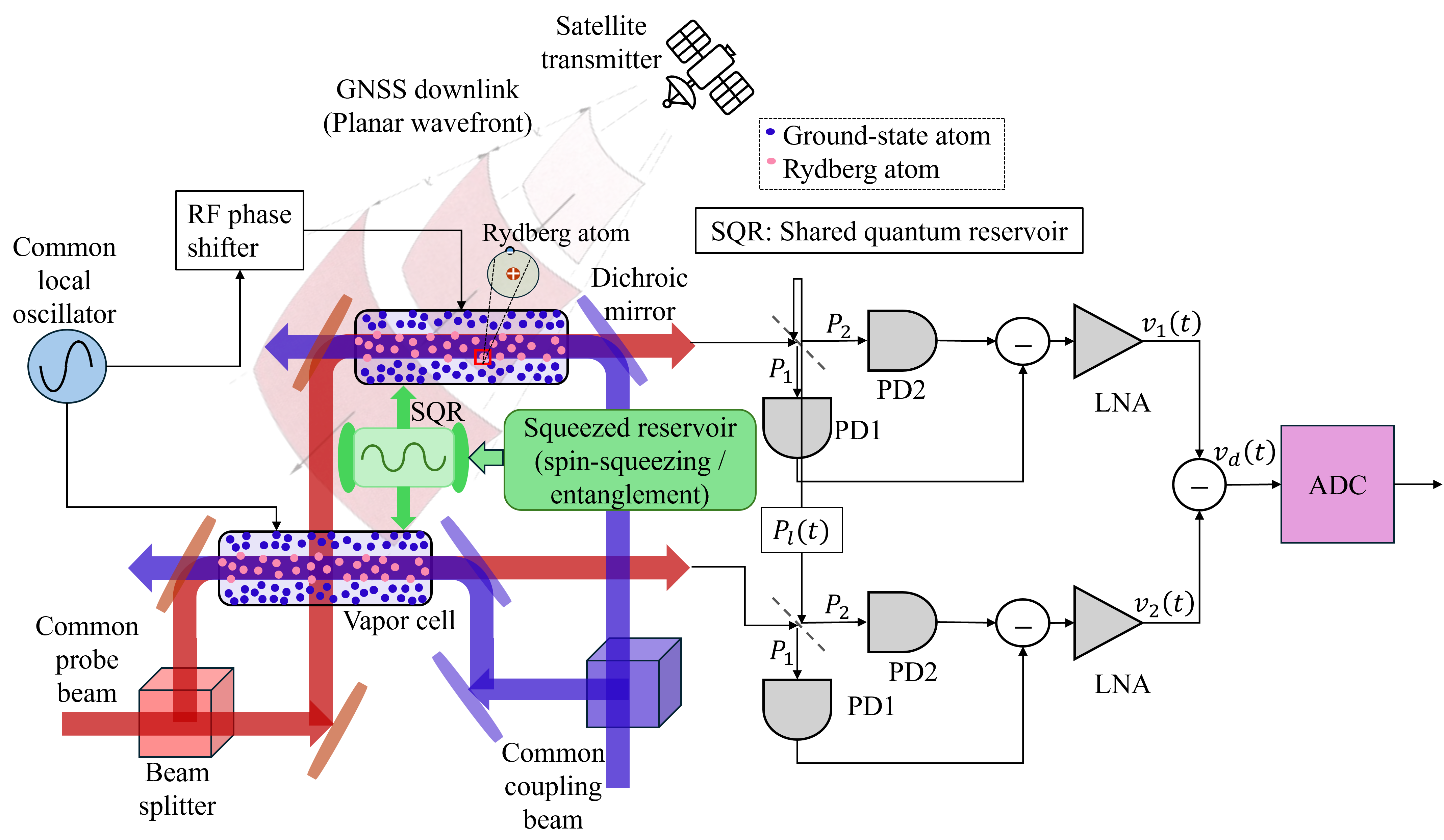}
\caption{DRAQR architecture; PD represents photodetector.}
\label{fig:draqr}
\end{figure}
The DRAQR in Fig.~\ref{fig:draqr} comprises two alkali-atom vapor cells (e.g., Cs or Rb) sharing counter-propagating probe and coupling lasers that establish EIT. The probe $\{\Omega_p,\omega_p,\Delta_p\}$ drives $|1\rangle\leftrightarrow|2\rangle$, and the coupling field $\{\Omega_c,\omega_c,\Delta_c\}$ drives $|2\rangle\leftrightarrow|3\rangle$, where $\Omega$, $\omega$, and $\Delta$ represent the corresponding Rabi, angular, and detuning frequencies. Both cells use common laser sources, while the incident RF field drives $|3\rangle\leftrightarrow|4\rangle$, modulating the transmitted probe's amplitude and phase.

As shown in Fig.~\ref{fig:draqr}, a cavity-mediated two-mode squeezed reservoir dissipatively prepares squeezed states in separated atomic ensembles, controlling their quantum fluctuations. This directs the received RF signal to the antisymmetric (differential) mode while correlating selected fluctuations in the common mode for rejection by differential detection. Inter-ensemble correlations depend on the reservoir coupling rate $\Gamma_R$, atomic decoherence rate $\Gamma_a$, and reservoir-to-atom transfer efficiency $\eta_R$, while the phase-controlled LO independently sets the relative phase of the RF responses. 

We assume closely spaced cells located at $\mathbf{r}_1$ and $\mathbf{r}_2$ experience the same RF field under far-field planar-wave propagation beyond $2L_{\rm Tx}^2/\lambda_x$, where $L_{\rm Tx}$ is the transmitter's largest dimension and $\lambda_x=c/f_x$, yielding $P_{b1}=P_{b2}=P_b$. Unlike conventional RAQRs, based on incident-RF-induced ATS, the LO here is coherently combined with the received RF field at both cells before atomic interaction. 
\section{Phase-Balanced DRAQR}\label{pbDraqr}
Consider a GNSS transmitter generating a binary sequence $\mathbf{b}=\{b_1, \cdots, b_n\}$. For $M$-QAM, the complex constellation is represented as $a_n=I_n+\iota Q_n$, $\mathcal{A}=\{a_1, \cdots, a_M\}$, and $\mathbb{E}[|a_n|^2]=1$. The interpolated signal is $s_b(t)=\sqrt{P_s}\sum_{n=-\infty}^{\infty}a_ng(t-nT_s)$, where $P_s$ is the average symbol power, $T_s$ is the symbol duration, and $g(t)$ is the pulse shaping filter. Assuming ideal interpolation, we have $g(t)=\sinc(Wt)$, where $W$ is the sampling rate and $\sinc(t)=\frac{\sin(\pi t)}{\pi t}$. The upconverted passband counterpart is given by
\begin{align}
\begin{split}
s(t)=\sqrt{2}\Re\left(s_b(t)\exp(\iota 2\pi f_x t)\right),
\end{split}
\end{align}
where $f_x$ is the carrier frequency. The equivalent baseband signal per symbol duration is $s_b(t)=\sqrt{P_s}e^{\iota\theta_s}$.
\subsection{Single-cell Signal Reception -- The Benchmark}
Consider a linear time-varying wireless communication channel $h(\tau, t)=\sum_i a_i(t)\delta[\tau-\tau_i(t)]$, with $a_i(t)$ and $\tau_i(t)$ denoting the gain and delay for the $i$th multipath component. The received basedband signal $x_b(m)$ is given as
\begin{align}\label{sigbb}
\begin{split}
x_b(t)=\sqrt{A_e}\sum_ia_i(t)e^{-\iota (2\pi f_x\tau_i(t)-\mathbf{K}_x^\top\mathbf{r})}s_b[t-\tau_i(t)],
\end{split}
\end{align}
where $\mathbf{K}_x$ is the propagation vector, with $A_e$ and $\mathbf{r}$ denoting the effective aperture and position of the RAQR, respectively. 
%
%
The RF signal incident on the RAQR is given as
\begin{align}
\begin{split}
x(t)=&\sqrt{2P_x}\cos(2\pi f_x t+\theta_x)=\sqrt{2}\Re\left\{x_b(t)e^{\iota 2\pi f_x t}\right\},
\end{split}
\end{align}
where $P_x$ and $\theta_x$ are the power and reference phase of the GNSS downlink signal, respectively. Considering a superheterodyne RAQR, the desired RF signal is given as $z(t)=x(t)+y_{\rm LO}(t)$, where $y(t)$ is the LO beam, given as
\begin{align*}
\small
\begin{split}
y_{\rm LO}(t)=&\sqrt{2P_{\rm LO}}\cos(2\pi f_{\rm LO} t+\theta_{\rm LO})=\sqrt{2}\Re\left\{y_{\rm LO}^{(\rm b)}(t)e^{\iota 2\pi f_{\rm LO} t}\right\},
\end{split}
\end{align*}
where $f_{\rm LO}$, $P_{\rm LO}$, and $\theta_{\rm LO}$ are the frequency, power, and phase, respectively, and $y_{\rm LO}^{(\rm b)}(t)=\sqrt{P_{\rm LO}}e^{\iota\theta_{\rm LO}}$ is the equivalent baseband signal, with $P_{\rm LO}$ and $\theta_{\rm LO}$ represent the power and phase of the LO signals. Consequently, the desired RF signal $z(t)$ is written as
\begin{align}
\begin{split}
z(t)= \sqrt{2P_z}\cos(2\pi f_{\rm LO} t+\theta_{\rm LO}),
\end{split}
\end{align}
where $P_z$ is the power of $z(t)$. 
%
%
The probe beam at the access area of the atomic vapor cell is given as
\begin{align}\label{probepb}
\begin{split}
y_p(t)=\sqrt{2P_p}\cos(2\pi f_p t+\phi_p)=\sqrt{2}\Re\left\{y_p^{(\rm b)}e^{\iota 2\pi f_p t}\right\},
\end{split}
\end{align}
where $P_p$, $f_p$, and $\phi_p$ are the power, frequency, and phase of the input probe beam, respectively. Furthermore, $y_p^{(\rm b)}=\sqrt{P_p}e^{\iota\phi_p}$ is the equivalent baseband signal of the passband probe beam in Eq.~\eqref{probepb}. The amplitude, $U_p'(\Omega_{\rm RF})$, and phase, $\phi_p'(\Omega_{\rm RF})$ of the output probe beam are given as
\begin{align}
\begin{split}
U_p'(\Omega_{\rm RF})=&U_p\exp(-\pi d\Im\left\{\chi(\Omega_{\rm RF})\right\}/\lambda_p)\\ 
\phi_p'(\Omega_{\rm RF})=&\phi_p+\pi d\Re\left\{\chi(\Omega_{\rm RF})\right\}/\lambda_p,
\end{split}
\end{align}
where $\lambda_p$ is the wavelength of the probe laser, $d$ is the length of the vapor cell, and $\chi(\Omega_{\rm RF})$ is the susceptibility of the atomic vapor cell \cite{mandal2026ber}. 
%
%
The output probe beam is given as 
\begin{align}
\begin{split}
y_{\rm op}(\Omega_{\rm RF}, t)=\sqrt{2}\Re\left\{y_{\rm op}^{(\rm b)}(\Omega_{\rm RF},t)e^{\iota 2\pi f_p t}\right\},
\end{split}
\end{align}
and the baseband probe beam signal $y_{\rm op}^{(\rm b)}(\Omega_{\rm RF},t)$ is given as
\begin{align}
\begin{split}
y_{\rm op}^{(\rm b)}(\Omega_{\rm RF},t)=\sqrt{P_{\rm op}(\Omega_{\rm RF})}e^{\iota\phi_p}e^{\iota\frac{\pi d}{\lambda_p}\Re\{\chi(\Omega_{\rm RF}\}},
\end{split}
\end{align}
where $P_{\rm op}(\Omega_{\rm RF})$ is the power of the output probe beam. Considering a balanced coherent optical detection, the local optical beam amplitude is expressed as
\begin{align}
\small
\begin{split}
y_l(t)=&\sqrt{2P_l}\cos(2\pi f_l t+\phi_l)=\sqrt{2}\Re\left\{y_l^{(\rm b)}(t)e^{\iota 2\pi f_l t}\right\},
\end{split}
\end{align}
where $P_l$ is the power of the local optical beam and $y_l^{(\rm b)}(t)=\sqrt{P_l}e^{\iota\phi_l}$ is the equivalent baseband waveform. 
%
%
For a load resistance $R_L$ and LNA gain $G$, the output AC voltage is \cite{mandal2026ber}
\begin{align}\label{dcac}
\begin{split}
\tilde{V}(t)=&2\alpha\sqrt{GP_lP_{\rm op}(\Omega_{\rm LO})}\kappa(\Omega_{\rm LO})\cos\varphi(\Omega_{\rm LO})\\
&\times U_x\cos(2\pi f_\delta t+\theta_\delta)=\sqrt{2}\Re\left\{\tilde{V}_b(t)e^{\iota 2\pi f_\delta t}\right\},
\end{split}
\end{align}
where $f_{\delta}=f_x-f_{\rm LO}$, $\alpha=\frac{\eta_1 q}{2\pi\hbar f_p}$, $\eta_1$ denotes the quantum efficiency, $q$ is the electron charge, $\kappa=\frac{\pi d\mu_{34}}{\lambda_p \hbar}|\chi'(\Omega_{\rm LO})|$, $\varphi(\Omega_{\rm LO})=\phi_l-\phi'_p(\Omega_{\rm LO})+\psi_p(\Omega_{\rm LO})$ with $\chi'(\Omega_{\rm LO})$ given in \cite{GongTCOM2026}, 
 $\psi_p(\Omega_{\rm LO})=\frac{\pi}{2}-\arctan\left(\frac{\Im\{\chi'(\Omega_{\rm LO})\}}{\Re\{\chi'(\Omega_{\rm LO})\}}\right)$, $\tilde{V}_b(t)=\sqrt{\varrho}\Phi x_b(t)$, $\varrho=4\alpha^2GP_lP_{\rm op}(\Omega_{\rm LO})\kappa^2(\Omega_{\rm LO})$, and $\Phi=\cos\varphi(\Omega_{\rm LO})e^{-\iota\theta_{\rm LO}}$. 
%
%
Sampling Eq.~\eqref{dcac} at $1/W$, we have
\begin{align}\label{vb}
\begin{split}
\tilde{V}_b(m)=\sqrt{\varrho}\Phi x_b(m).
\end{split}
\end{align}
Next, upon sampling the received signal at intervals of $1/W$, Eq.~\eqref{sigbb} is written as
\begin{align}\label{rxsignonoise}
\begin{split}
x_b(m)=e^{-\iota\mathbf{k}_x^\top\mathbf{r}}\sqrt{A_e}\sum_{l}h_l(m)s_b(m-l),
\end{split}
\end{align}
where $h_l(m)$ is the equivalent discrete-time channel impulse response \cite{GongTCOM2026}. Therefore, using Eq.~\eqref{rxsignonoise} and considering an AWGN $w(m)$, Eq.~\eqref{vb} is rewritten as
\begin{align}\label{rxsig}
\begin{split}
\tilde{V}_b(m)=\sqrt{\varrho}\Phi e^{-\iota\mathbf{k}_x^{\top}\mathbf{r}}\sum_lh_l(m)s_b(m-l)+w(m).
\end{split}
\end{align}
The total receiver noise $w(m)$ is modeled as \cite{GongTCOM2026}
\begin{align}
\begin{split}
\mathbb{E}[\|w(m)\|^2]=\sigma_w^2=\frac{1}{2}\left(\sigma^2_{\rm QPN}+\sigma^2_{\rm PSN}+\sigma^2_{\rm ITN}\right),
\end{split}
\end{align}
where $\sigma^2_{\rm QPN}$, $\sigma^2_{\rm PSN}$, and $\sigma^2_{\rm ITN}$ are QPN, PSN, and ITN powers.
\subsection{Differential Signal Reception}
Considering narrowband fading and similar fading across both Rydberg cells, i.e., $h_1(m)\stackrel{d}{=}h_2(m)=h(m)$, the received baseband signals, as per Eq.~\eqref{rxsig}, are given as
\begin{align}
\begin{split}
\tilde{V}_{b1}(m)=&\sqrt{\varrho_1}\Phi_1 e^{-\iota\mathbf{k}_x^\top\mathbf{r}_1}h_1(m)s_b(m)+w_1(m)\\
\tilde{V}_{b2}(m)=&\sqrt{\varrho_2}\Phi_2e^{-\iota\mathbf{k}_x^\top\mathbf{r}_2}h_2(m)s_b(m)+w_2(m).
\end{split}
\end{align}
Thus, $v_d(m)=\tilde{V}_{b1}(m)-\tilde{V}_{b2}(m)$, is obtained as
\begin{align}
\begin{split}
v_d(m)=&g_d\sqrt{\varrho}e^{-\iota\mathbf{k}_x^\top\mathbf{r}_1}h(m)s_b(m)+\Delta w,
\end{split}
\end{align}
where $g_d=\sqrt{\varrho_1}\Phi_1 -\sqrt{\varrho_2}\Phi_2 e^{-\iota\mathbf{k}_x^\top\Delta\mathbf{r}}$, $\Delta w=w_1(m)-w_2(m)$, and $\Delta\mathbf{r}=\mathbf{r}_2-\mathbf{r}_1$. The differential noise is written as
\begin{align*}
\begin{split}
\Delta w=&(w_{\rm QPN}^{(1)}-w_{\rm QPN}^{(2)})+(w_{\rm PSN}^{(1)}-w_{\rm PSN}^{(2)})+(w_{\rm ITN}^{(1)}-w_{\rm ITN}^{(2)}),
\end{split}
\end{align*}
where the superscripts denote the cell number. Therefore, the differential noise power is given as $P_{\Delta w}=\mathbb{E}[\|\Delta w(m)\|^2]$
\begin{align}\label{noiseExp}
\small
\begin{split}
P_{\Delta w}=\frac{1}{2}\sum_{i=1}^2\sigma^2_{w,i}-2\cov(w_1,w_2),
\end{split}
\end{align}
where the subscript $i$ represents the corresponding noise for the $i$th Rydberg cell. Notably, for a noise type $x$, the covariance can be computed as $\cov(w_x^{(1)}, w_x^{(2)})=\rho_x\sigma_x^{(1)}\sigma_x^{(2)}$, where $\rho_x$ is the correlation in noise type $x$ between the two cells and $\sigma^2$ is the noise power. Thus, $\cov(w_1,w_2)$ is simplified as
\begin{align}
\small
\begin{split}
2\cov(w_1,w_2)=&\rho_{\rm QPN}\sigma_{{\rm QPN},1}\sigma_{{\rm QPN},2}+\rho_{\rm PSN}\sigma_{{\rm PSN},1}\sigma_{{\rm PSN},2}\\
&+\rho_{\rm ITN}\sigma_{{\rm ITN},1}\sigma_{{\rm ITN},2}.
\end{split}
\end{align}
The SNR for the differential reception is written as
\begin{align}
\begin{split}
\text{SNR}_d=\frac{|g_d|^2|h(m)|^2P_b}{P_{\Delta w}},
\end{split}
\end{align}
where $P_b=\mathbb{E}[|s_b(m)|^2]$. The SNR gain $G_{\rm SNR}$ between single-cell ($\text{SNR}_s$) and differential signal reception is given as
\begin{align}
\small
\begin{split}
G_{\rm SNR}=&\frac{\text{SNR}_d}{\text{SNR}_s}=\frac{|g_d|^2\sigma_w^2}{\varrho\Phi^2_1 P_{\Delta w}}.
\end{split}
\end{align}
The minimum detectable power is calculated with the following guarantee on outage probability
\begin{align}\label{outProb}
\begin{split}
\text{Pr}\left\{|h(m)|^2\leq \gamma_{\rm th}\frac{P_{\Delta w}}{|g_d|^2P_b}\right\} \leq P_{\rm out}^{(\max)},
\end{split}
\end{align}
where $P_{\rm out}^{(\max)}$ is the maximum outage probability for successful signal detection. Eq.~\eqref{outProb} can be simplified considering various distributions for the wireless fading channel. However, a generic bound is obtained using Chebyshev's inequality
\begin{align}
\begin{split}
\text{Pr}\left\{|h(m)|^2\leq \gamma_{\rm th}\frac{P_{\Delta w}}{|g_d|^2P_b}\right\}\leq \frac{\sigma_h^2}{\left(\mu_h-\gamma_{\rm th}\frac{P_{\Delta w}}{|g_d|^2P_b}\right)^2},
\end{split}
\end{align}
where $\mu_h=\mathbb{E}[|h(m)|^2]$ and $\sigma_h^2=\mathbb{E}[(|h(m)|^2-\mu_h)^2]$. Imposing probability limit, the minimum detectable power is
\begin{align}
\begin{split}
P_b^{(\min)}\approx\frac{\gamma_{\rm th}P_{\Delta w}}{|g_d|^2}\sqrt{P_{\rm out}^{(\max)}}\left(\mu_h\sqrt{P_{\rm out}^{(\max)}}-\sigma_h\right)^{-1}.
\end{split}
\end{align}
\subsection{Modeling of Noise Correlation $\rho_{\rm QPN}$, $\rho_{\rm PSN}$, $\rho_{\rm ITN}$}\label{corrMod}
\textbf{Derivation of $\rho_{\rm QPN}$:} Let the collective transverse spin fluctuations of cells $1$ and $2$ be represented by bosonic operators $\hat{b}_1$ and $\hat{b}_2$. 
Define the collective modes
\begin{align}
\begin{split}
\hat{b}_{+}=\frac{\hat{b}_1+\hat{b}_2}{\sqrt{2}},\quad \hat{b}_{-}=\frac{\hat{b}_1-\hat{b}_2}{\sqrt{2}},
\end{split}
\end{align}
where $\hat{b}_+$ is the common mode and $\hat{b}_-$ is the differential mode. The GPS signal is deliberately placed in $b_-$. 
%
%
Let the shared quantum reservoir have two modes, $\hat{b}_{R1}$ and $\hat{b}_{R2}$, associated with the vapor cells. Define a quadrature and its conjugate as
\begin{align*}
\begin{split}
\hat{X}_{R,m}=\frac{\hat{b}_{Rm}+\hat{b}_{Rm}^\dagger}{\sqrt{2}}, \quad \hat{P}_{R,m}=\frac{\hat{b}_{Rm}-\hat{b}_{Rm}^\dagger}{\iota\sqrt{2}}, \; m\in\{1,2\}.
\end{split}
\end{align*}
We define the variance and covariance as $V_R=\var(\hat{X}_{R1})=\var(\hat{X}_{R2})$ and $C_R=\cov(\hat{X}_{R1},\hat{X}_{R2})$, respectively. For ideal two-mode squeezed vacuum with squeezing parameter $r$, we have $V_R=\frac{1}{2}\cosh 2r$ and $C_R=\frac{1}{2}\sinh 2r$.
%
%
Since DRAQR subtracts the two cell outputs, its variance is given as
\begin{align}
\footnotesize
\begin{split}
\var(X_{R,-})=\frac{1}{2}[\var(\hat{X}_{R1})+\var(\hat{X}_{R2})-2\cov(\hat{X}_{R1},\hat{X}_{R2})].
\end{split}
\end{align}
Considering a participation factor $\eta_R=\Gamma_R/(\Gamma_R+\Gamma_a)$, we use the decoherence model $\hat{X}_m=\sqrt{\eta_R}\hat{X}_{Rm}+\sqrt{1-\eta_R}\hat{X}_{{\rm vac},m}$, where $\hat{X}_{{\rm vac},m}$ represents vacuum noise. Then, 
\begin{align}
\begin{split}
V_X=\var(\hat{X}_m)=\frac{1}{2}[1-\eta_R+\eta_R\cosh 2r].
\end{split}
\end{align}
Since independent vacuum contributions have zero correlation, so $C_X=\eta_RC_R=\frac{\eta_R}{2}\sinh 2r$. Thus, $\rho_{\rm QPN}$ is defined as 
\begin{align}
\small
\begin{split}
\rho_{\rm QPN}=\frac{C_X}{V_X}=&\frac{\eta_R\sinh 2r}{1-\eta_R+\eta_R\cosh 2r}= \frac{\Gamma_R\sinh 2r}{\Gamma_a+\Gamma_R\cosh 2r}.
\end{split}
\end{align}

\textbf{Derivation of $\rho_{\rm PSN}$:} Let $\hat{a}_1$ and $\hat{a}_2$ denote the annihilation operators of the two optical modes associated with the probe fields of cells $1$ and $2$, respectively. Considering a two-mode squeezed vacuum with squeezing parameter $r_O$, the corresponding amplitude quadratures are defined as
\begin{align}
\begin{split}
\hat{X}_{O,m}=(\hat{a}_m+\hat{a}_m^\dagger)/\sqrt{2}, \qquad m\in\{1,2\}.
\end{split}
\end{align}
%
%
For the individual optical modes, the quadrature variance is
\begin{align}
\begin{split}
V_{\rm PSN}^{(R)}&=\frac{1}{2}\left\langle\hat{a}_m\hat{a}_m^\dagger+\hat{a}_m^\dagger\hat{a}_m+\hat{a}_m^2+\hat{a}_m^{\dagger2}\right\rangle.
\end{split}
\end{align}
For a two-mode squeezed vacuum, the single-mode anomalous moments satisfy $\langle\hat{a}_m^2\rangle=\langle\hat{a}_m^{\dagger2}\rangle=0$, while $\langle\hat{a}_m^\dagger\hat{a}_m\rangle=\sinh^2 r_O$ and $\langle\hat{a}_m\hat{a}_m^\dagger\rangle=1+\sinh^2 r_O$, resulting in $V_{\rm PSN}^{(R)}=0.5\cosh(2r_O)$.
%
%
The inter-quadrature covariance is given as
\begin{align*}
\small
\begin{split}
C_{\rm PSN}^{(R)}&=\left\langle\hat{X}_{O,1}\hat{X}_{O,2}\right\rangle=\frac{1}{2}\left[\langle\hat{a}_1\hat{a}_2\rangle+\langle\hat{a}_1\hat{a}_2^\dagger\rangle+\langle\hat{a}_1^\dagger\hat{a}_2\rangle+
\langle\hat{a}_1^\dagger\hat{a}_2^\dagger\rangle\right].
\end{split}
\end{align*}
For the two-mode squeezed vacuum, $\langle\hat{a}_1\hat{a}_2\rangle=M_R^{(O)}$, $\langle\hat{a}_1^\dagger\hat{a}_2^\dagger\rangle=M_R^{(O)}$, $\langle\hat{a}_1\hat{a}_2^\dagger\rangle=0$, $\langle\hat{a}_1^\dagger\hat{a}_2\rangle=0$, and $C_{\rm PSN}^{(R)}=\frac{1}{2}\sinh(2r_O)$. 
%
%
The differential optical quadrature for the proposed DRAQR is $\hat{X}_{O,-}=\frac{\hat{X}_{O,1}-\hat{X}_{O,2}}{\sqrt{2}}$ with its variance being
\begin{align*}
\small
\begin{split}
V_{O,-}&=\var(\hat{X}_{O,-})=\frac{1}{2}\left[V_{\rm PSN}^{(R)}+V_{\rm PSN}^{(R)}-2C_{\rm PSN}^{(R)}\right]\\
&=V_{\rm PSN}^{(R)}-C_{\rm PSN}^{(R)}=\frac{1}{2}\left[\cosh(2r_O)-\sinh(2r_O)\right]=\frac{1}{2}e^{-2r_O}.
\end{split}
\end{align*}
Let $\eta_O\in[0,1]$ denote the total optical transmission and detection efficiency. The effect of this loss can be represented by a beam-splitter transformation $\hat{a}_m^{({\rm det})}=\sqrt{\eta_O}\hat{a}_m+\sqrt{1-\eta_O}\hat{v}_m$, where $\hat{v}_m$ is an independent vacuum mode introduced by the loss channel. The corresponding quadrature is
\begin{align}
\begin{split}
\hat{X}_{O,m}^{({\rm det})}&=\sqrt{\eta_O}\hat{X}_{O,m}+\sqrt{1-\eta_O}\hat{X}_{{\rm vac},m}.
\end{split}
\end{align}
The vacuum quadrature satisfies $\var(\hat{X}_{{\rm vac},m})=\frac{1}{2}$ and $\cov(\hat{X}_{{\rm vac},1},\hat{X}_{{\rm vac},2})=0$. Thus, the variance measured in either optical detection channel is
\begin{align}
\begin{split}
V_{\rm PSN}&=\var\left(\hat{X}_{O,m}^{({\rm det})}\right)=\frac{1}{2}\left[1-\eta_O+\eta_O\cosh(2r_O)\right].
\end{split}
\end{align}
The cross-covariance between the optical channels is
\begin{align}
\small
\begin{split}
C_{\rm PSN}&=\cov\left(\hat{X}_{O,1}^{({\rm det})},\hat{X}_{O,2}^{({\rm det})}\right)=\eta_OC_{\rm PSN}^{(R)}.
\end{split}
\end{align}
Hence, $C_{\rm PSN}=\frac{\eta_O}{2}\sinh(2r_O)$ resulting in $\rho_{\rm PSN}$ given as
\begin{align}
\begin{split}
\rho_{\rm PSN}&=\frac{C_{\rm PSN}}{V_{\rm PSN}}=\frac{\eta_O\sinh(2r_O)}{1-\eta_O+\eta_O\cosh(2r_O)}.
\end{split}
\end{align}
\begin{note}
Assuming independent detector chains, $\rho_{\rm ITN}\to 0$.
\end{note}
\subsection{Cell Matching}
For a planar wave, the propagation constant can be denoted as $\mathbf{k}_x=\frac{2\pi}{\lambda_x}\hat{\mathbf{k}}_x$, where $\hat{\mathbf{k}}_x$ is an unit vector in the direction of propagation and $\lambda_x=\frac{c}{f_x}$ is the carrier wavelength. Therefore, the path-dependent phase offset can be given as
\begin{align}
\begin{split}
\mathbf{k}_x^\top\Delta\mathbf{r}=2\pi\lambda_x^{-1}\hat{\mathbf{k}}_x^\top\Delta\mathbf{r}=2\pi f_xc^{-1}\hat{\mathbf{k}}_x^\top\Delta\mathbf{r}.
\end{split}
\end{align}
This phase can be ignored for cells that are illuminated by wavelengths larger than the distance $\Delta r$ between them, leading to $\mathbf{k}_x^\top\Delta\mathbf{r}\to 0$. Thus, cell matching is obtained by controlling the phase of the LO to obtain $\sqrt{\varrho_1}\Phi_1=-\sqrt{\varrho_2}\Phi_2$, i.e.,
\begin{align}
\begin{split}
\alpha_1\kappa_1\sqrt{P_{\rm op}^{(1)}}e^{-\iota\theta_{\rm LO}^{(1)}}=-\alpha_2\kappa_2\sqrt{P_{\rm op}^{(2)}}e^{-\iota\theta_{\rm LO}^{(2)}},
\end{split}
\end{align}
where $\theta_{\rm LO}^{(1)}$ and $\theta_{\rm LO}^{(2)}$ are the phases of the LO signal fed to cells $1$ and $2$, respectively. Under perfect term-wise matching, we have $\alpha_1=\alpha_2=\alpha$, $\kappa_1=\kappa_2=\kappa$, and $P_{\rm op}^{(1)}=P_{\rm op}^{(2)}=P_{\rm op}$. Thus, the final cell matching is obtained by maintaining a phase relation $\theta_{\rm LO}^{(2)}-\theta_{\rm LO}^{(1)}=(2k+1)\pi$, where $k\in\mathbb{Z}$.

However, for cases where the path-difference--induced phase shift cannot be ignored, i.e., $\lambda_x$ is comparable to $\Delta\mathbf{r}$, cell matching is obtained by adjusting the LO phase to obtain
\begin{align}
\begin{split}
\alpha_1\kappa_1\sqrt{P_{\rm op}^{(1)}}e^{-\iota\theta_{\rm LO}^{(1)}}=-\alpha_2\kappa_2\sqrt{P_{\rm op}^{(2)}}e^{-\iota(\theta_{\rm LO}^{(2)}+\mathbf{k}_x^\top\Delta\mathbf{r})}.
\end{split}
\end{align}
After term-matching for phase-balancing, we obtain $\theta_{\rm LO}^{(2)}-\theta_{\rm LO}^{(1)}=(2k+1)\pi-\mathbf{k}_x^\top\Delta\mathbf{r}$. Considering the noise powers from \cite{GongTCOM2026} and considering the correlation modeling from Subsection \ref{corrMod}, we get Eq.~\eqref{noiseComb}, where $m_{\rm QPN}=\varrho c\epsilon_0\hbar^2N^{-1}B$, $m_{\rm PSN}=2.77q\alpha G\pi c\epsilon_0F_p^2B$, $m_{\rm ITN}=k_BGB$, $N=\Upsilon N_0 V$ is the number of atoms in the $V$ atomic volume at the population rate $\Upsilon$, and $k_B$ is the Boltzmann constant. 
%
%
\begin{table*}[t!]
\begin{align}\label{noiseComb}
\begin{split}
P_{\Delta w}&=\frac{1}{2}\left(\cos^2\varphi_1(\Omega_{\rm LO})\frac{\Gamma_{\rm EIT}^{(1)}}{\mu_{34,1}^2}+\cos^2\varphi_2(\Omega_{\rm LO})\frac{\Gamma_{\rm EIT}^{(2)}}{\mu_{34,2}^2}-2\rho_{\rm QPN}\cos\varphi_1(\Omega_{\rm LO})\cos\varphi_2(\Omega_{\rm LO})\frac{\sqrt{\Gamma_{\rm EIT}^{(1)}\Gamma_{\rm EIT}^{(2)}}}{\mu_{34,1}\mu_{34,2}}\right)m_{\rm QPN}\\
&+\frac{1}{2}\left(2|U_l|^2+|U'_{p,1}(\Omega_{\rm LO})|^2+|U'_{p,2}(\Omega_{\rm LO})|^2-2\rho_{\rm PSN}\sqrt{|U_l|^2+|U'_{p,1}(\Omega_{\rm LO})|^2}\sqrt{|U_l|^2+|U'_{p,2}(\Omega_{\rm LO})|^2}\right)m_{\rm PSN}+\frac{1}{2}(T_1+T_2)m_{\rm ITN}
\end{split}
\end{align}
\hrulefill
\end{table*}
%

\textbf{Perfect matching:} Considering $\cos^2\varphi_1(\Omega_{\rm LO})\Gamma_{\rm EIT}^{(1)}\mu_{34,1}^{-2}= \cos^2\varphi_2(\Omega_{\rm LO})\Gamma_{\rm EIT}^{(2)}\mu_{34,2}^{-2}$, $|U'_{p,1}(\Omega_{\rm LO})|=|U'_{p,2}(\Omega_{\rm LO})|$, $\rho_{\rm QPN}=\rho_{\rm PSN}=1$, and $T_1=T_2=T_0$, the noise power reduces to $P_{\Delta w}=Gk_BT_0B$. Thus, $g_d=2\sqrt{\varrho}\Phi$, leading to
\begin{align}
\small
\begin{split}
G_{\rm SNR}=2\left[1+(\sigma^2_{\rm QPN}+\sigma^2_{\rm PSN})/\sigma^2_{\rm ITN}\right].
\end{split}
\end{align}
Therefore, under perfect matching, $P_b^{(\min)}$ is obtained as
\begin{align}
\begin{split}
P_b^{(\min)}\approx\frac{\gamma_{\rm th}Gk_BT_0B}{4\varrho\Phi^2}\frac{\sqrt{P_{\rm out}^{(\max)}}}{\mu_h\sqrt{P_{\rm out}^{(\max)}}-\sigma_h}.
\end{split}
\end{align}
%
\section{Results and Discussions}\label{res}

As a representative weak signal, we consider a GPS L1 C/A signal reception. The optimal settings used for the Rydberg vapor cells are as suggested in \cite{mandal2026ber}. MC simulations assume a coherent hard-decision detection, perfect channel knowledge with over $150,000$ data runs per result. The generated MC data points are shown with markers in corresponding graphs. The RF antenna reference considers the L1 band operation, noise temperature of $300$ K, and antenna gain of $5.5$ dBi. 
The list of simulation parameters used for these results are taken from \cite{GongTCOM2026}. 
For comparison, a conventional GNSS receiver \cite{9726792} and a high-sensitive (HS) GNSS receiver \cite{7914663} are considered. 
\subsection{Inter-cell Noise Correlation}
\begin{figure}[t!]
\centering
\subfigure[]{\includegraphics[width=1.71in]{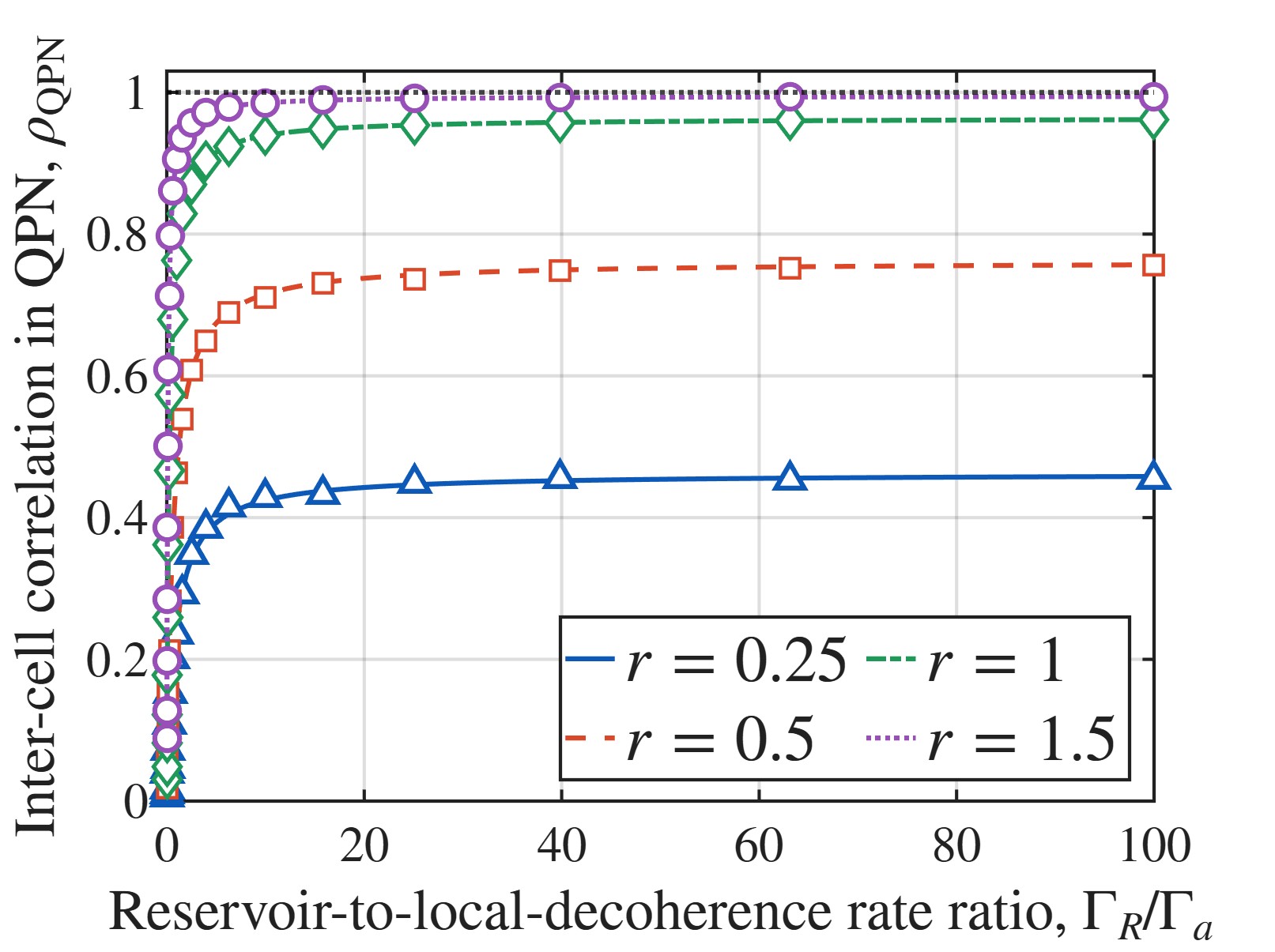}}
\subfigure[]{\includegraphics[width=1.71in]{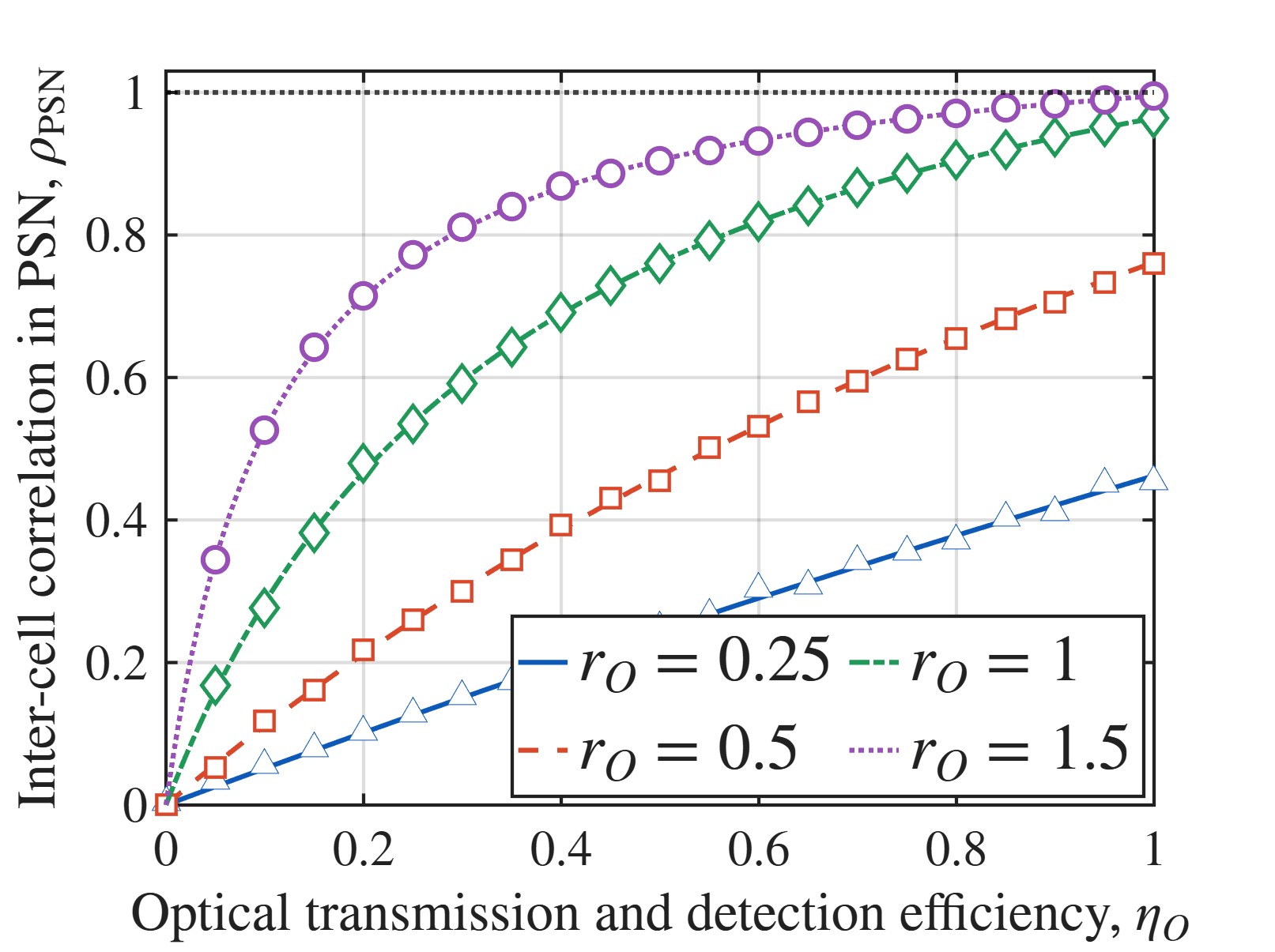}}
\vspace{-3mm}
\caption{INC variation: (a) $\rho_{\rm QPN}$ vs. $\frac{\Gamma_{R}}{\Gamma_a}$ and (b) $\rho_{\rm PSN}$ vs. $\eta_O$.}
\label{fig:icc}
\end{figure}
Fig.~\ref{fig:icc} shows that stronger reservoir coupling and optical squeezing increase the correlations available to differential reception. In Fig.~\ref{fig:icc}(a), for $r=1.5$, $\rho_{\rm QPN}$ rises to approximately $0.90$ at $\Gamma_R/\Gamma_a=1$ and approaches $1$ by $\Gamma_R/\Gamma_a=100$. For the weaker squeezing case, $r=0.25$, it instead saturates near $0.46$. Thus, increasing $\Gamma_R/\Gamma_a$ is most useful before the curves saturate; after which additional coupling produces little further correlation. Fig.~\ref{fig:icc}(b) exhibits the role of optical transmission. At $\eta_O=0.8$, the plotted $\rho_{\rm PSN}$ values are approximately $0.38$, $0.66$, $0.90$, and $0.97$ for $r_O=0.25$, $0.5$, $1$, and $1.5$, respectively. This makes clear that squeezing can sustain strong PSN correlation despite non-ideal optical efficiency, while loss remains a limiting factor at weaker squeezing. 

\subsection{SNR Gain}
\begin{figure}[t!]
\centering
\subfigure[]{\includegraphics[width=1.71in]{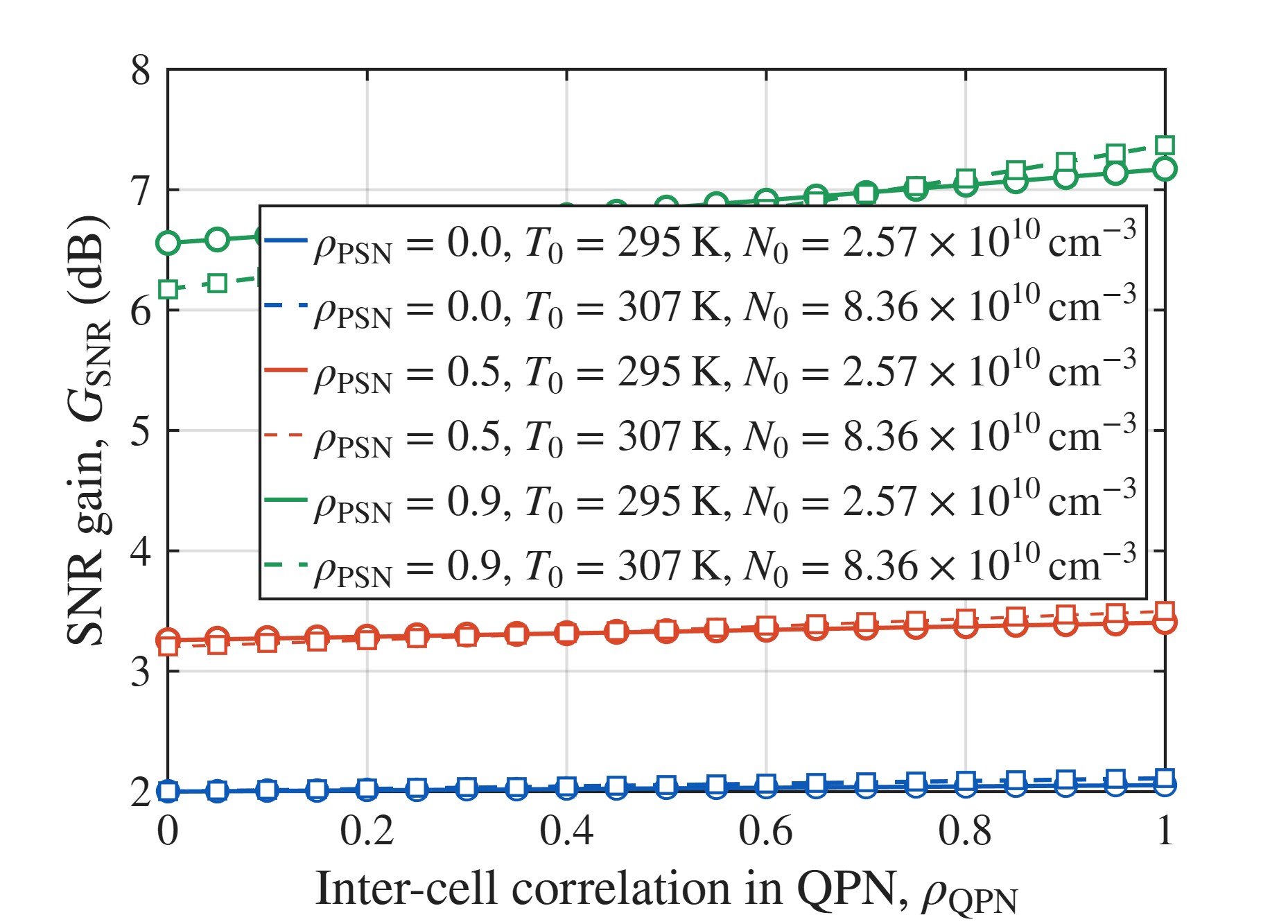}}
\subfigure[]{\includegraphics[width=1.71in]{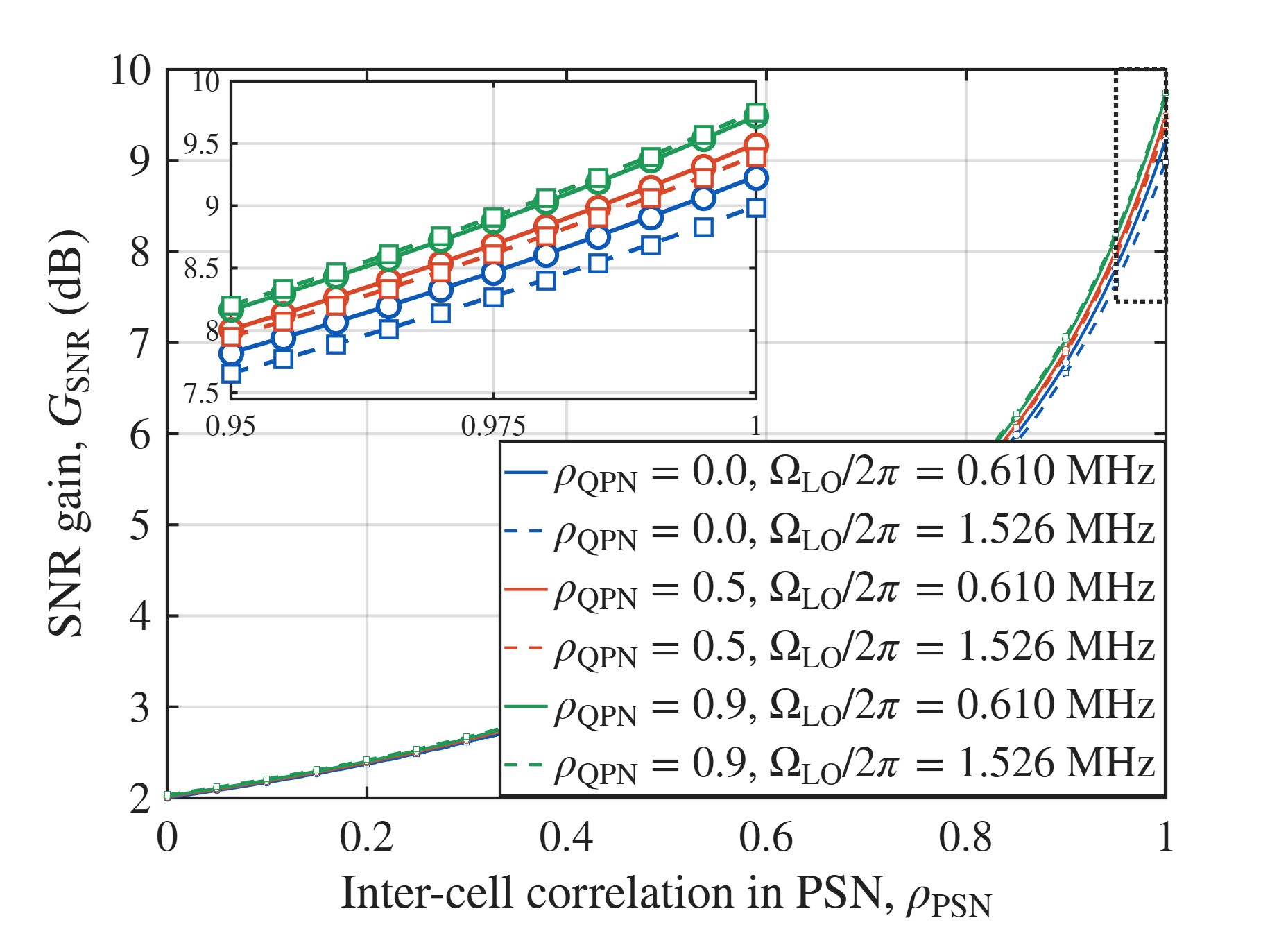}}
\vspace{-3mm}
\caption{SNR gain versus (a) $\rho_{\rm QPN}$ and (b) $\rho_{\rm PSN}$.}
\label{fig:snrg}
\end{figure}
The results shown in Fig.~\ref{fig:snrg}(a) demonstrate that increasing $\rho_{\rm QPN}$ produces a modest gain for fixed $\rho_{\rm PSN}$. The plotted curves rise by only a few tenths of a decibel over most of the correlation range. By contrast, changing $\rho_{\rm PSN}$ from $0$ to $0.9$ raises the gain from roughly $2$~dB to about $6.5$--$7.4$~dB, depending on the temperature--density pair. This difference is consistent with Table~I, where the reference PSN power is about $3.8$ times the ITN power and over $18$ times the QPN power. Fig.~\ref{fig:snrg}(b) reinforces this result as the gain here is seen to rise from $2$~dB at $\rho_{\rm PSN}=0$ to $9.5$~dB at $\rho_{\rm PSN}\to 1$. $\rho_{\rm QPN}$ and LO-frequency cases remain close over much of the range, with their separation becoming visible at high $\rho_{\rm PSN}$. The principal design opportunity shown here is therefore to suppress the dominant, correlated PSN contribution.

\subsection{Minimum Detectable Power}
\begin{figure}[t!]
\centering
\subfigure[]{\includegraphics[width=1.71in]{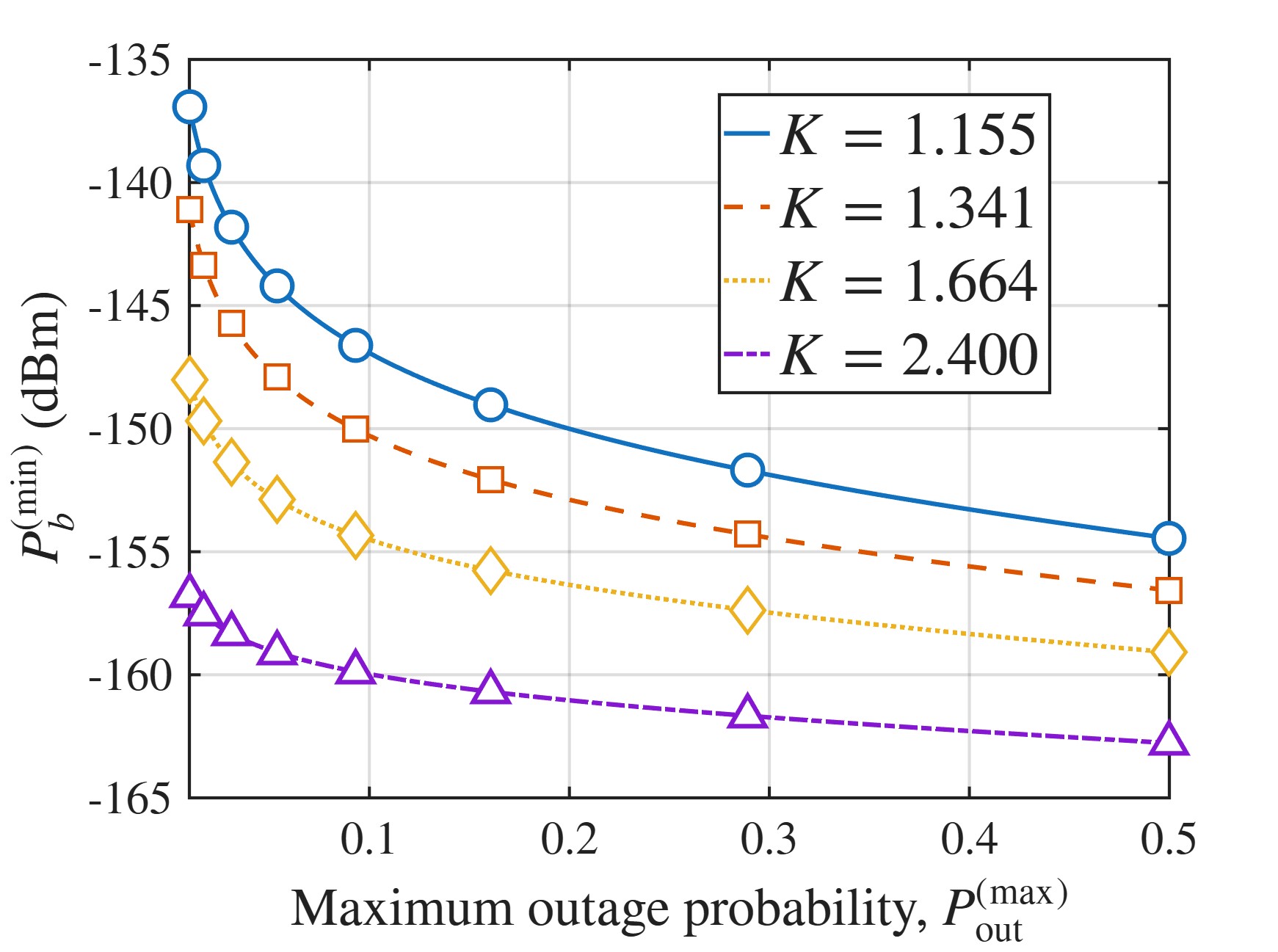}}
\subfigure[]{\includegraphics[width=1.71in]{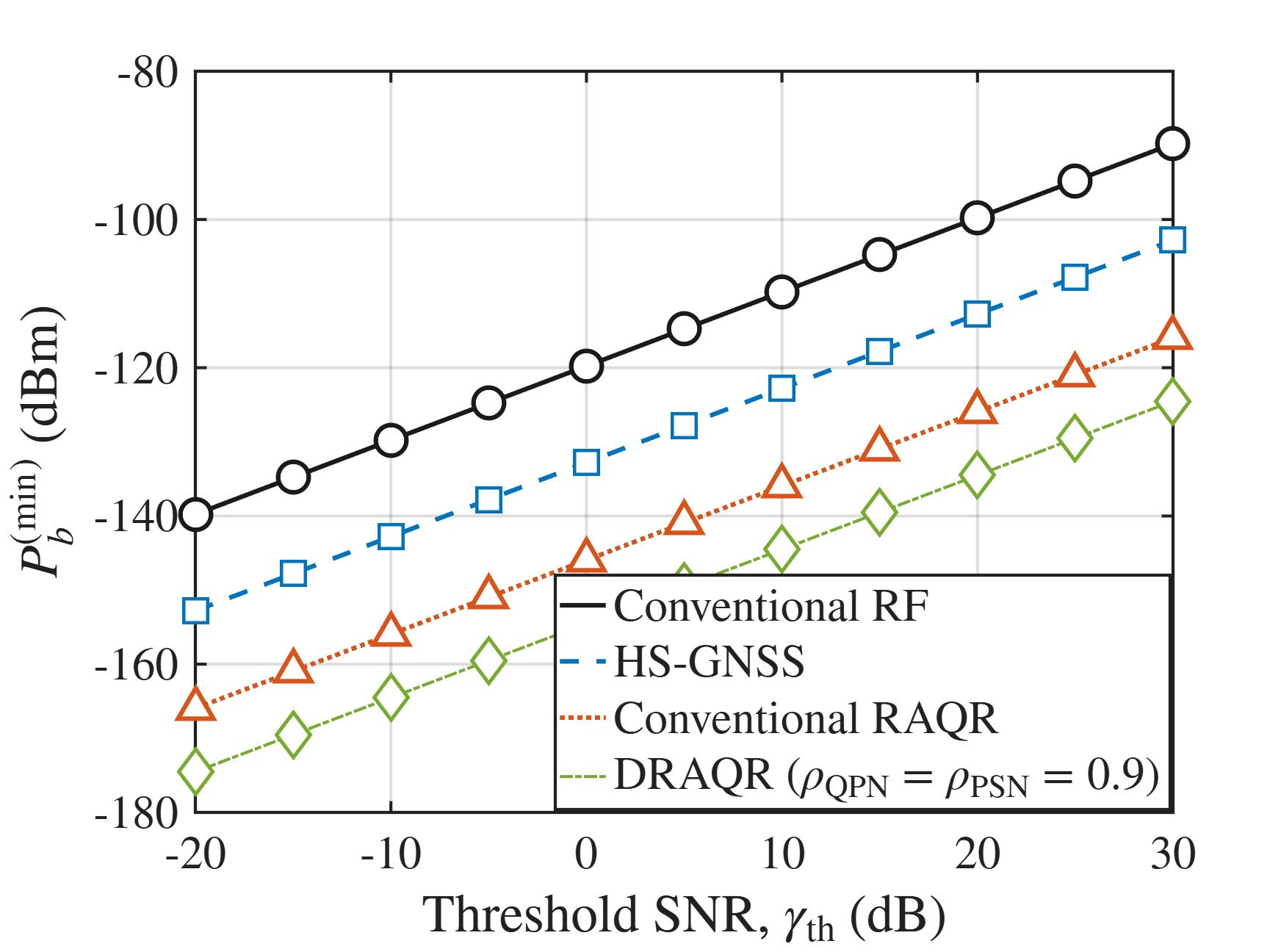}}
\vspace{-3mm}
\caption{Minimum power $P_b^{(\min)}$ versus (a) $P_{\rm out}^{(\max)}$ and (b) $\gamma_{\rm th}$.}
\label{fig:pmin}
\end{figure}
Fig.~\ref{fig:pmin}(a) shows the expected outage trade-off, where allowing a larger $P_{\rm out}^{(\max)}$ lowers the required received power because the
receiver can tolerate deeper fades. At $P_{\rm out}^{(\max)}\approx0.1$, the plotted minimum power falls from roughly $-146$~dBm for
$K=1.155$ to about $-160$~dBm for $K=2.4$, a reduction of approximately $14$~dB (about a 25-fold power reduction). At the
stricter $0.01$ outage level, the separation is closer to $20$~dB. These curves also confirm that channel statistics materially affect the power needed
for reliable recovery.

Fig.~\ref{fig:pmin}(b) isolates the receiver comparison: at $\gamma_{\rm th}=0$~dB, the plotted minimum powers are approximately $-120$~dBm for conventional GNSS receiver, $-133$~dBm for the $20$-ms coherent HS-GNSS case, $-146$~dBm for conventional RAQR, and $-154.5$~dBm for DRAQR. Thus, DRAQR requires about $8.5$~dB less power than a conventional RAQR (roughly seven times less) and about $34.5$~dB less than a conventional GNSS receiver. The curves are nearly parallel across the displayed threshold-SNR range, indicating that these receiver advantages persist as the required detection threshold changes.

\subsection{BER Performance}
\begin{figure}[t!]
\centering
\subfigure[]{\includegraphics[width=1.12in]{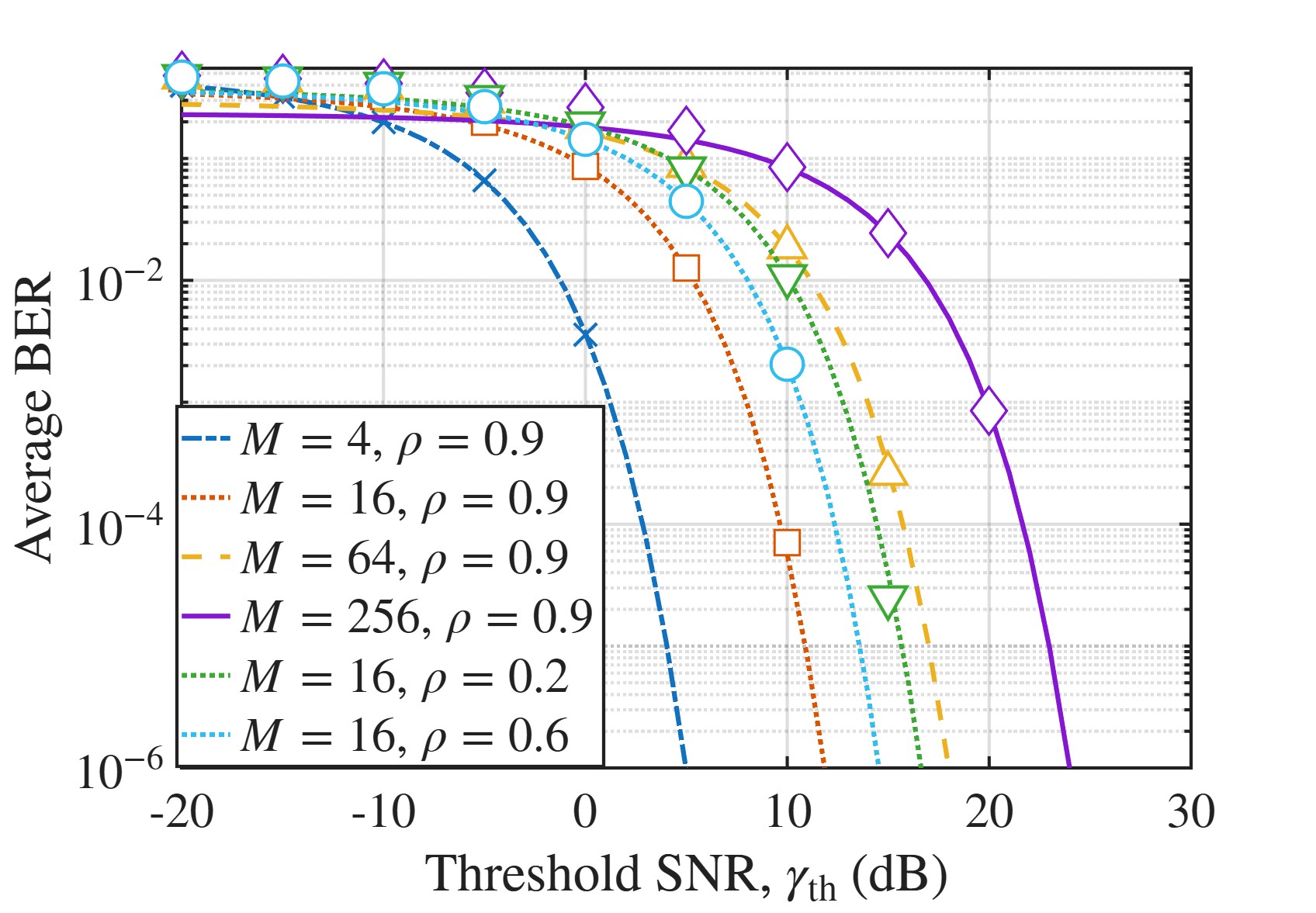}}
\subfigure[]{\includegraphics[width=1.12in]{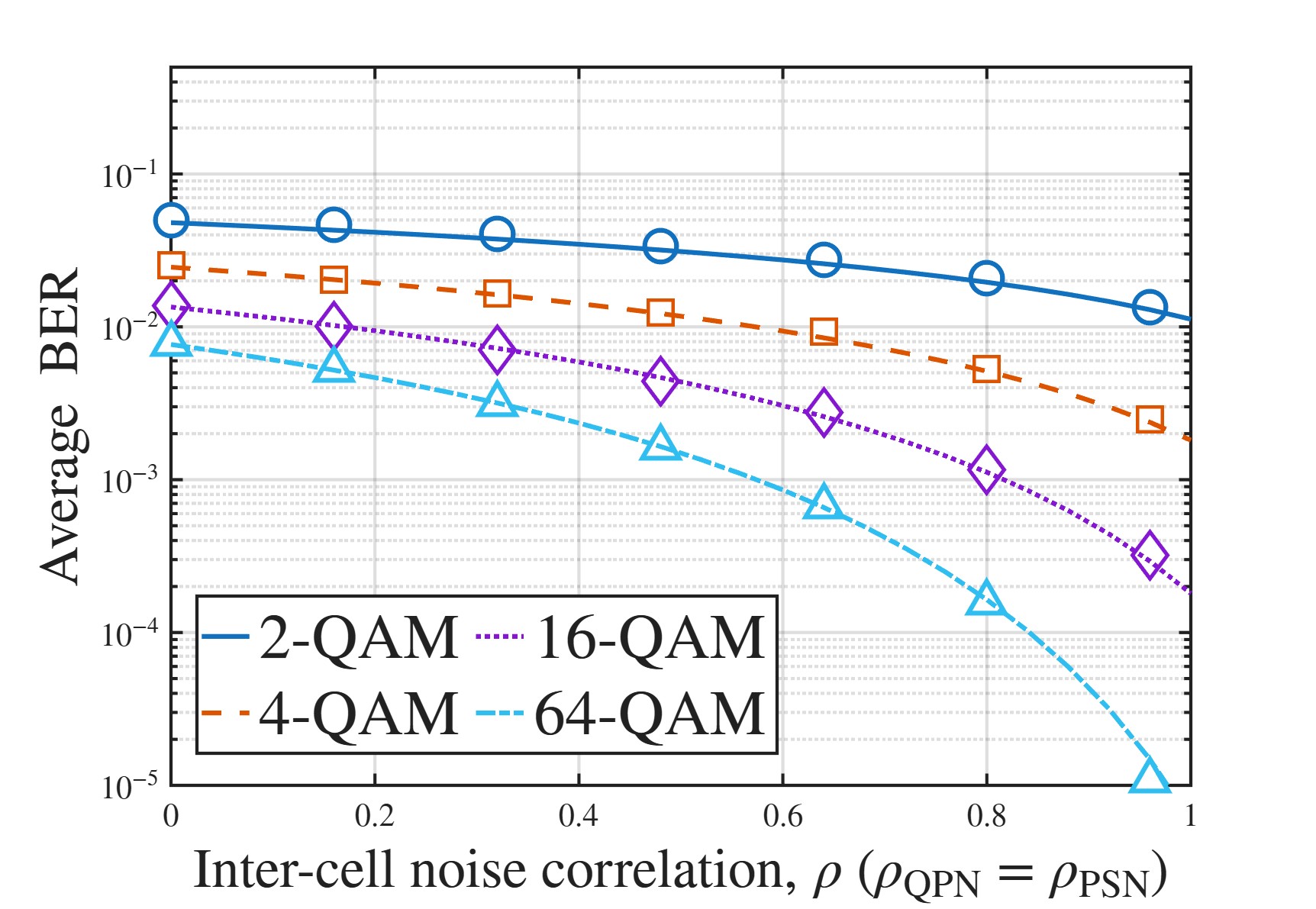}}
\subfigure[]{\includegraphics[width=1.12in]{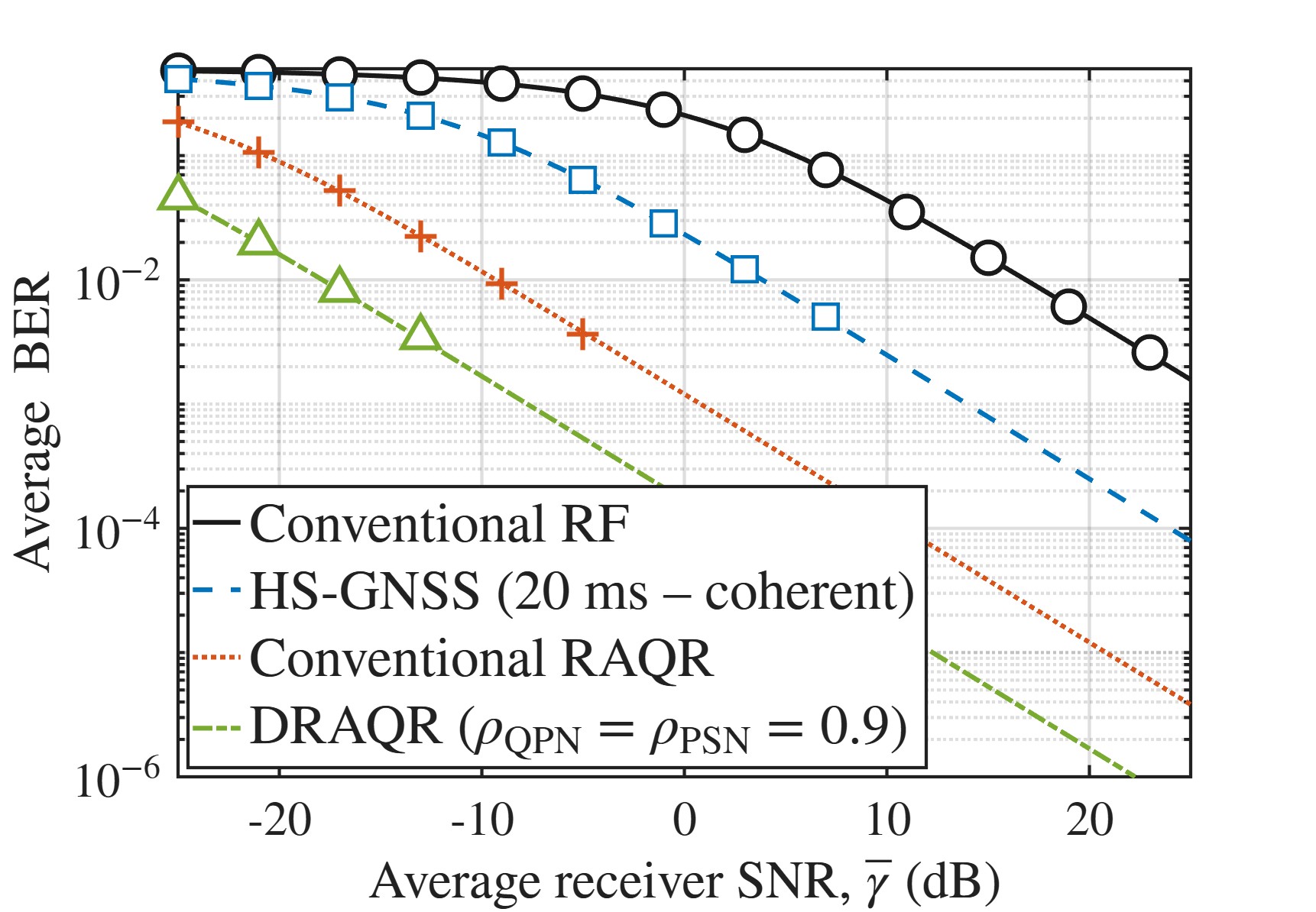}}
\vspace{-3mm}
\caption{BER vs. (a) $\gamma_{\rm th}$, (b) $\rho$, ($\rho_{\rm QPN}=\rho_{\rm PSN}$), and (c) $\bar{\gamma}$.}
\label{fig:berp}
\end{figure}
Here, we consider the conditional BER expression for a square Gray-coded $M$-QAM \cite{mandal2026ber}.
%
In Fig.~\ref{fig:berp}(a), higher threshold SNR reduces BER, whereas higher QAM order shifts the curves right. At BER $\approx10^{-3}$ and $\rho=0.9$, the required thresholds are approximately $-5$, $5$, $14$, and $22$~dB for $M=4$, $16$, $64$, and $256$, respectively. For $M=16$, increasing $\rho$ from $0.2$ to $0.9$ lowers the required threshold by $6$--$8$~dB. Fig.~\ref{fig:berp}(b) shows that 64-QAM BER decreases from $\approx8\times10^{-3}$ at $\rho=0$ to $1$--$2\times10^{-4}$ near $\rho=0.8$ ($40$ to $80$-fold reduction). In Fig.~\ref{fig:berp}(c), at BER $\approx10^{-3}$, DRAQR requires an average SNR of $\approx-4$~dB, versus $\approx+5$~dB for conventional RAQR and $+13$--$15$~dB for HS-GNSS; the conventional GNSS receiver does not reach this BER within the plotted range. These results agree with the minimum-power comparison, showing DRAQR's strongest BER advantage over conventional RAQR, conventional RF, and HS-GNSS receivers.

%
\section{Conclusion}\label{conc}
This work demonstrates the potential of the proposed DRAQR to improve SNR and BER by coordinating signal and noise modes across two vapor cells. LO phase control directed the desired response to the differential mode, while correlations in quantum projection noise (QPN) and photon shot noise (PSN) suppressed residual noise. Receiver-level gains depended on both atomic transduction and readout covariance, and not field sensitivity alone. Outage analysis further highlighted the dependence of required signal power on propagation statistics and detection criteria. Analytical results closely agree with Monte-Carlo simulations for the evaluated settings. However, these benefits depend on the assumed inter-cell covariances and phase balance. Future work intends to conduct measurements for QPN and PSN and assess the effects of phase errors, cell mismatch, and implementation losses.
%

%

\bibliography{IEEEabrv,IEEEref1}{}
\bibliographystyle{IEEEtran}

\end{document}